\documentclass[sigconf, nonacm]{acmart}

\newcommand\vldbpagestyle{plain}

\newcommand{\titlename}{Systems for Parallel and Distributed Large-Model Deep Learning Training}
\newcommand{\techreporteat}{}

\usepackage{hyperref,array,color,balance,multirow}
\usepackage{balance,float,url,amsfonts,alltt}
\usepackage{mathtools,rotating,amsmath}

\usepackage{etoolbox,listings}
\usepackage{bigstrut,morefloats,pbox}
\usepackage{graphicx,subfigure,xspace,verbatim,comment}
\usepackage{grffile}
\usepackage{ifpdf,fancyvrb}
\usepackage{algorithm}
\usepackage[noend]{algorithmic}
\usepackage{booktabs}
\usepackage{lipsum}  
\usepackage[all]{nowidow}
\usepackage{multirow}
\usepackage{pifont}
\usepackage{xcolor}

\newcommand{\eat}[1]{}
 
 \definecolor{mygreen}{HTML}{3EC300}

\usepackage{scalerel}

\newtoggle{TR}

\begin{document}\sloppy
\title{\titlename}

\author{Kabir Nagrecha}
\affiliation{%
\institution{University of California, San Diego}
\city{La Jolla}
\country{USA}
}
\email{knagrech@ucsd.edu}

\begin{abstract}
Deep learning (DL) has transformed applications in a variety of domains, including computer vision, natural language processing, and tabular data analysis. The search for improved DL model accuracy has led practitioners to explore
increasingly large neural architectures, with some recent Transformer models spanning hundreds of billions of learnable parameters. These designs have introduced new scale-driven systems challenges for the DL space, such as memory 
bottlenecks, poor runtime efficiency, and high costs of model development. Efforts to address these issues have explored techniques such as parallelization of neural architectures, spilling data across the memory hierarchy, and memory-efficient data representations. This survey will explore the large-model training systems landscape, highlighting key challenges and the various techniques that have been used to address them.
\end{abstract}
\maketitle

\pagestyle{\vldbpagestyle}
\techreporteat{

\vspace{.3cm}
}

\section{Introduction}
\label{sec:intro}
Over the last several years, deep learning (DL) has cemented itself as a critical component of a variety of high-value applications.
Computer vision (CV), natural language processing (NLP), and recommender systems now rely heavily on DL architectures to 
enable predictive modeling and analysis. The popularity of DL has led to the emergence of a new area, known as ``Systems for DL''
that studies the infrastructural and computational challenges introduced by the use of DL. 

\begin{figure}[th!]
\centering
	\includegraphics[keepaspectratio=true, width=\linewidth]{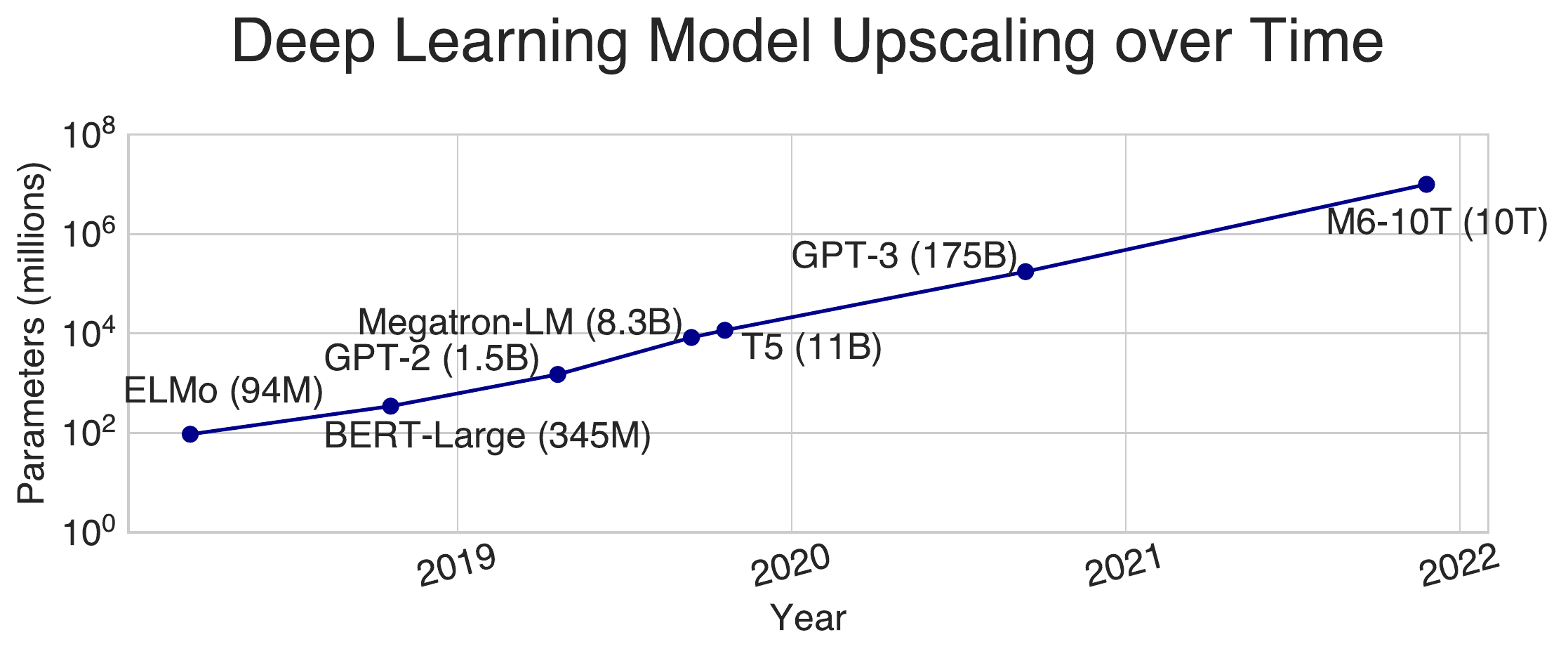}
	\caption{Illustration of how state-of-the-art DL Transformer models have grown over time. Y-axis (model parameter counts) presented in logarithmic scale.}
	\label{fig:scaling}
\end{figure}

Recent developments in DL practice have introduced a pressing new challenge of \textit{model scale} in systems for DL research. Practitioners
have begun to explore the use of very large neural architecture graphs for DL models, with some containing billions and 
even \textit{trillions} of trainable parameters! Key examples include the NLP Transformer models BERT-Large~\cite{bert2018}, GPT-3~\cite{gpt2020}
and Meta's deep learning recommender model (DLRM)~\cite{dlrm2019}. The sheer sizes of these models have introduced critical challenges in three key areas.

\begin{enumerate}
\item \textbf{Memory Scalability.} Standard DL training typically holds a model's parameters on the memory of an accelerator (e.g. a GPU) and uses sampled data to compute gradient updates for each parameter. For a very large model, the space required to hold the parameters, intermediate computations, and gradient updates will typically exceed the relatively limited memory of an accelerator. High-end consumer-grade GPUs such as the Tesla V100~\cite{teslav100} have 16-32GB of on-device memory, but a large-scale DLRM might require hundreds of gigabytes of space.
\item \textbf{Performance.} Increased parameter counts are typically correlated with higher execution times. In addition, complex model architectures tend to require large datasets to saturate the model's learning capacity --- GPT-3 was trained on 300B tokens, for example~\cite{gpt2020} and the open-source BLOOM language model was trained on 366B~\cite{bloom2022}. Training such models can require weeks, or even months~\cite{dlrm2019,bloom2022}. As such, optimizations that can improve execution performance are highly beneficial
for developers of large-scale DL models.
\item \textbf{Cost of Training.} Standard solutions to the previous two challenges typically involve parallelizing storage or execution across multiple devices. However, this approach can drive up compute costs significantly. BLOOM was trained on 416 A100 GPUs, and Megatron-LM used 512~\cite{megatronlmblog2020}. This is not practical for most practitioners, particularly when these GPUs need to be reserved for a training period of weeks or even months. Reproducing BLOOM's training procedure on AWS
would cost 5.5 million dollars. Note that this does not even account for the added costs of model selection, wherein multiple models are trained to evaluate the best hyperparameter settings and configurations~\cite{kumar2021cerebro}.
\end{enumerate}

As practitioners continue to push the bounds of model scale, it becomes increasingly necessary that these challenges are addressed to enable further development in the large-model DL space. As a result, various systems and techniques have been developed tackling one or another of these problems. Some directions include rematerialization~\cite{checkpointing2016}, data spilling/CPU offloading
~\cite{zero2019, zero2021,hydra2021,mpms2021,swapadvisor2021,l2l2020}, pipeline/model parallelism~\cite{gpipe2018,pipedream2018,terapipe2021,torchgpipe2020,megatronlmgpuscaling2021}, and hybrid parallelism~\cite{flexflow2018,alpa2022,hydra2021,mpms2021,gshard2020}. These subjects, falling under the general umbrella of ``large-model training techniques'', has become a key focus for researchers across industry and academia, but the sheer volume of work in the space has made this topic difficult to navigate. This paper will provide a comprehensive review of the current state of the large-model DL training systems space, along with an assessment of future directions of growth and development in the area.

A few high-level, short surveys on this area have recently been released~\cite{surveylarge2022,lowmemory2019} but they are not comprehensive, and do not address key topics such as model selection, hybrid parallelism, and the intersection of techniques. This paper will address these areas and also go into further depth on state-of-the-art techniques.

Note that this survey will not cover \textit{graph neural network} training as that class of architectures encounters substantially different problems from other DL models, and would require an entirely separate survey to explore in-depth~\cite{gnnsurvey2021}. It will also not cover ``model reduction'' techniques (e.g. model distillation) as these fall outside of the scope of training and go into model modification~\cite{wang2021gradient,polino2018model,gou2021knowledge}. For brevity's sake, we will also not cover Mixture-of-Expert models, which introduce differing (and complex) performance models.

This paper is organized as follows: Section~\ref{sec:background} goes over some necessary background and terminology. Section~\ref{sec:large_model} describes large-model architectures and some key examples. Section~\ref{sec:mlsys} explores the landscape of large-model DL training systems. Finally, Section~\ref{sec:conclusion} concludes the survey and explores possible future directions and emerging challenges.

\section{Background and Preliminaries}
\label{sec:background}
We start with some background on key technical concepts and notation needed for the rest of this paper.
\subsection{Deep Learning}
Deep learning (DL) refers to a technique for approximating patterns within data by applying gradient descent to optimize the parameters of a directed graph of operators. In this paper, we refer to the DL model graph as a \textit{model}, or a \textit{neural computational graph}. Feedforward networks are a particular type of computational graph wherein inputs are continuously fed forward through a chain of operators, each of which is known as a layer. Most large-scale model architectures fall under this class.

Common DL operators include \textit{matrix multiplies}, \textit{differentiable activation functions}, and \textit{embedding table lookups} (functionally equivalent to matrix multiplies with one-hot vectors). The first two are used for directly transforming data inputs while the latter is used for mapping a categorical input to some vector (e.g. translating an application user into a vector representing that user in a latent space). A DL model might use all of these operators together in a complex structure, typically known as a deep neural network (DNN).

The \textit{training} of a model involves an optimization procedure called stochastic gradient descent (SGD). It samples \textit{mini-batches} (disjoint subsets) of data from the training dataset, then runs a \textit{forward} pass through the model graph. The forward pass transforms the input data features into an output target, then calculates a \textit{loss} value, or error, relative to some ground truth \textit{label}. This loss value is used to run a \textit{backward} pass, also known as backpropagation. Backpropagation refers to the process of repeatedly applying the derivative chain rule to calculate gradient-based updates for each model parameter that will minimize the output loss. Note that in order for this chain process to work, intermediate outputs, also known as \textit{activations}, between operators \textit{must be saved}. \textit{Inference} applications, wherein the model parameters are frozen and deployed for prediction, can discard intermediate outputs when transforming inputs, but training has to save the intermediates (which can bloat memory costs).

A full pass of mini-batches through the entire dataset  (i.e., enough subsets have been sampled that the model has seen all the data) is called an \textit{epoch}. Training a model typically requires many epochs of SGD. Popular DL tools implement many variants of SGD (e.g., Adam, AdaGrad, RMSProp, etc.), known as \textit{optimizers} but their data access patterns are identical. Some optimizers maintain their own parameters and state which are updated during execution

\begin{figure*}[th!]
\centering
	\includegraphics[keepaspectratio=true, width=0.9\linewidth]{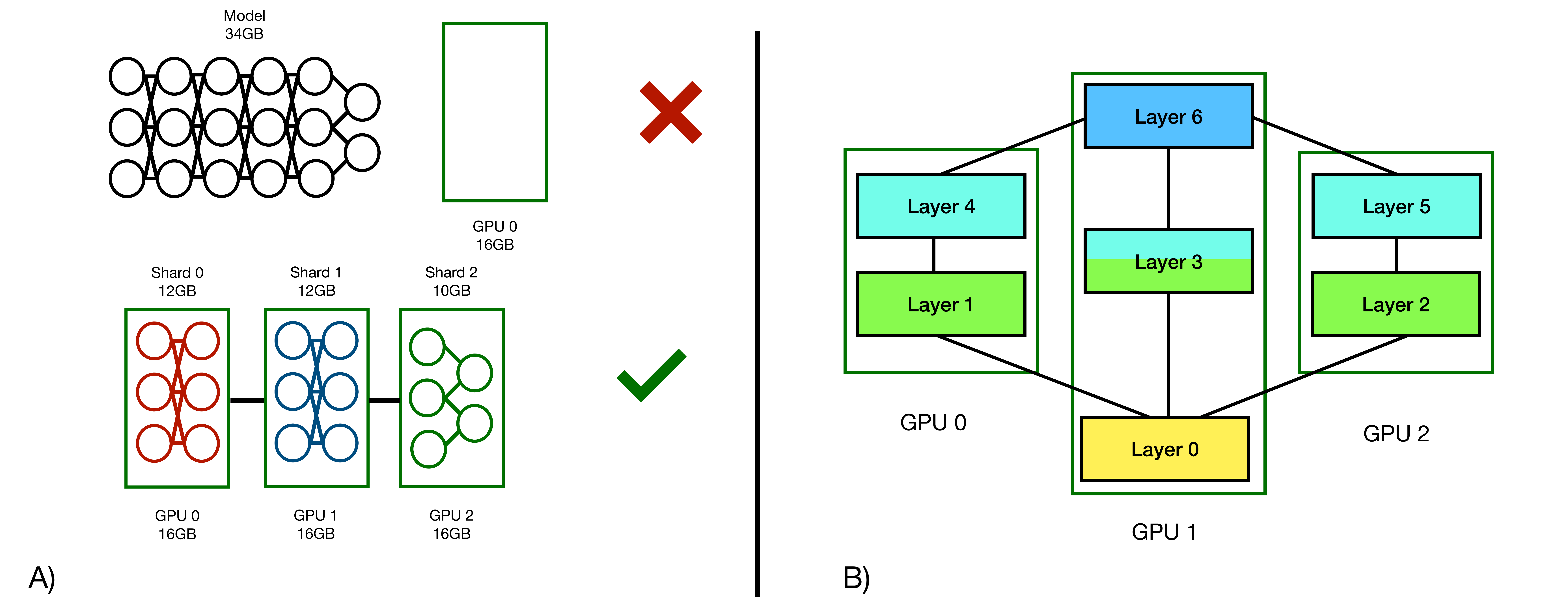}
	\caption{A) An illustration of how a large feedforward network that does not fit into a single GPU could be model-parallelized over three GPUs to enable execution. Note that execution has not been sped up --- there is no parallel execution, only partitioned memory demands. B) An illustration of a performant model parallel sharding strategy over three GPUs. Parallely executing layers are denoted with shared colors. This strategy exploits existing parallel execution opportunities within the operator graph --- this may not always be possible if the graph is more sequential in nature.}
	\label{fig:model_parallel_combined}
\end{figure*}

The size of a DL model can be measured through the number of parameters and/or its memory footprint. The memory footprint of a model is generally correlated with the number of parameters. In order to compute a full mini-batch-pass, intermediate data must be materialized between layers, and gradients will also need to be held in memory as they are chained backwards during backpropagation. As such, an upper bound of model memory footprint can be made based on the sum of its parameters' memory costs, its intermediates' memory costs, and its gradients' memory costs. This upper bound is typically a high overestimate, since most training frameworks will automatically discard intermediates and gradients as soon as they are no longer needed.

\subsection{Deep Learning Accelerators}
Matrix multiplies are typically the most common and computationally intensive operators within DL model graphs. Other operators exist (e.g. activation functions, embedding table lookups) but tend to be less computationally intense. The general matrix multiply operation, or GEMM, is a well studied target for parallelization~\cite{blislab2016}. As such, using hardware that can exploit this opportunity effectively is a common practice among DL model developers.

\textbf{GPUs} offer stronger parallelization capabilities versus CPUs for simple operations such as the GEMM. In particular, Nvidia GPUs support the cuDNN library~\cite{cudnn}, which implements low-level DL operators tuned for parallelization on GPU threads. However, GPUs must operate on data stored within on-accelerator memory, which is far more limited than standard system memory (DRAM). A state-of-the-art consumer GPU typically has less than 40GB of on-device memory, while DRAM capacity on a training machine might easily exceed 512GB. 

\textbf{TPUs}, or tensor processing units, are a new type of accelerator developed by Google specifically for DL. TPU cores are organized to support parallelism in GEMM operations and experimental evaluations show that it can achieve up to 10X faster performance than GPUs on some model architectures~\cite{tpubenchmark2019}. Unfortunately, TPU access is still very limited --- practitioners can only use one through Google's Cloud Platform. As such, this survey will primarily focus on systems targeting GPU execution.

\subsection{Parallelization Techniques}\label{sec:parallelization}
A variety of techniques exist for parallelizing DL training. Each offers different advantages or drawbacks, which tend to vary depending on the workload. Understanding the data access and communication patterns of each strategy is critical to evaluating them in the context of large-model training.

\textbf{Model parallelism} refers to the technique of partitioning, or \textit{sharding}, a neural architecture graph into subgraphs, and assigning each subgraph, or \textit{model shard}, to a different device. In a feedforward network, these shards might refer to groups of stacked layers.

\eat{
\begin{figure}[th!]
\centering
	\includegraphics[keepaspectratio=true, width=0.9\linewidth]{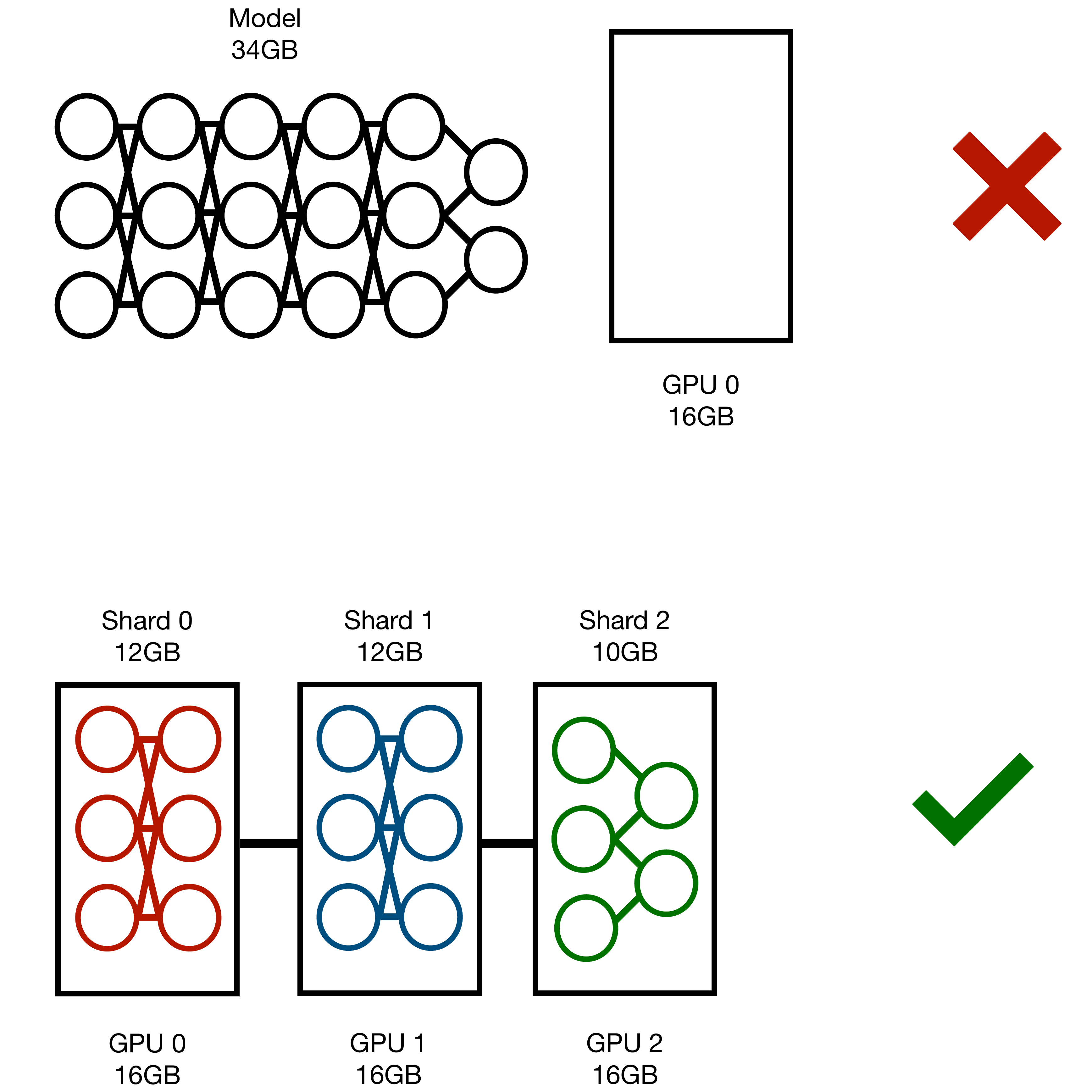}
	\caption{An illustration of how a large feedforward network that does not fit into a single GPU could be model-parallelized over three GPUs to enable execution. Note that execution has not been sped up --- there is no parallel execution, only partitioned memory demands.}
	\label{fig:model_parallel_feedforward}
\end{figure}

\begin{figure}[th!]
\centering
	\includegraphics[keepaspectratio=true, width=0.9\linewidth]{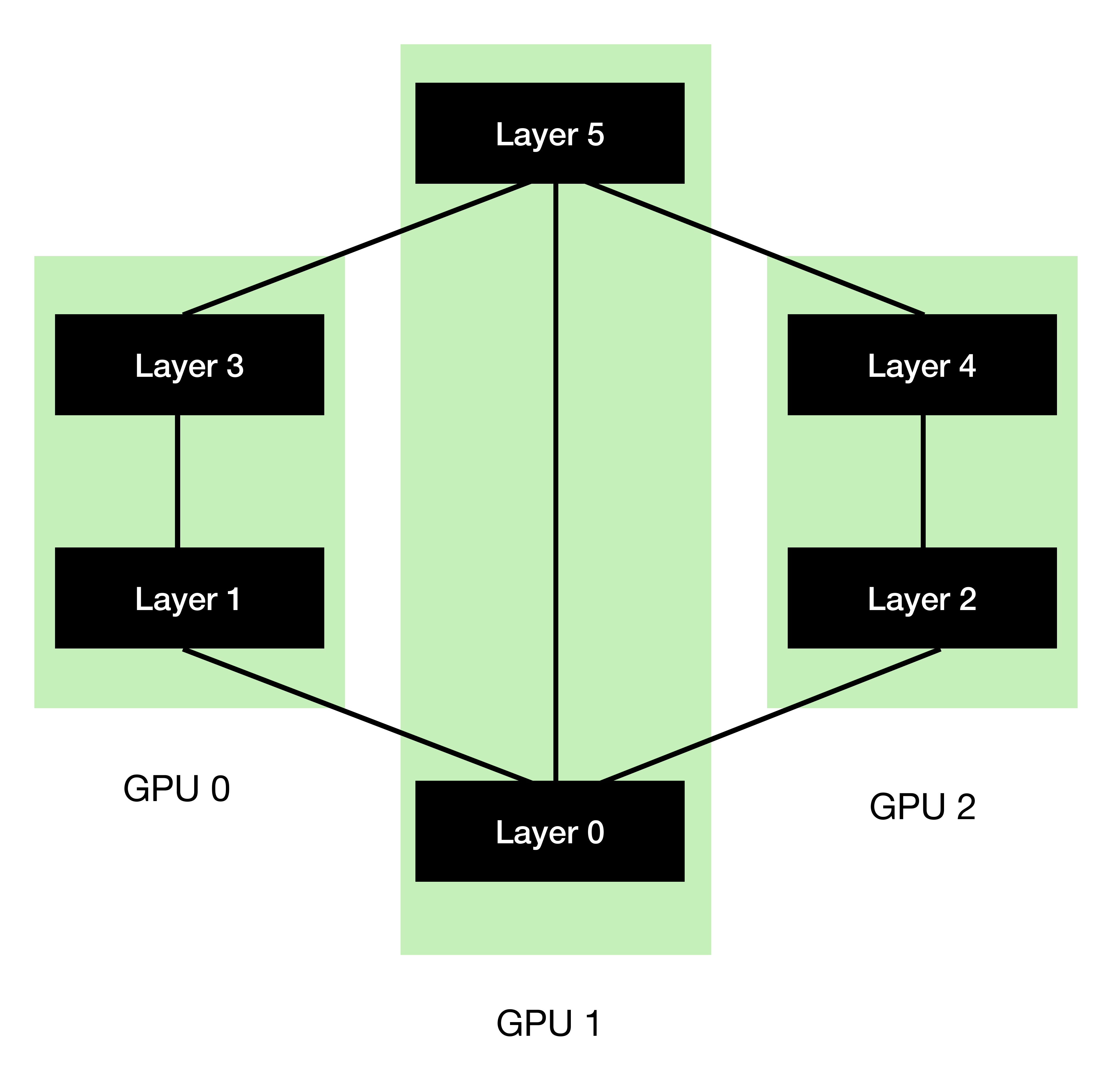}
	\caption{An illustration of a performant model parallel sharding strategy over three GPUs. This strategy exploits existing parallel execution opportunities within the operator graph --- this may not always be possible if the graph is more sequential in nature.}
	\label{fig:model_parallel_partitioned}
\end{figure}
}

The speedup potential of model parallelism depends largely on the architecture and the sharding strategy. Sequential model sharding on a feedforward network, of the sort illustrated in Figure~\ref{fig:model_parallel_combined}A) will offer no scope for parallel execution, instead inducing a dependency graph between accelerators. This sharding strategy is still popular for sequential architectures (e.g. Transformers), however, as it can distribute memory demands across multiple devices and is fairly simple to setup. In other cases, the neural computational graph offers natural opportunities for inter-operator parallelism. Figure~\ref{fig:model_parallel_combined}B) illustrates.

\begin{figure*}[th!]
\centering
	\includegraphics[keepaspectratio=true, width=0.9\linewidth]{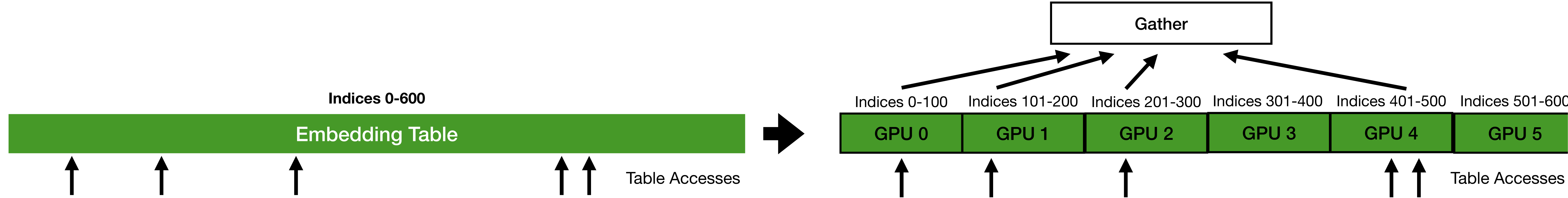}
	\caption{An embedding table parallelized over 6 GPUs. Each GPU receives a different subset of the table's indices.}
	\label{fig:embedding_table_parallel}
\end{figure*}

Another approach is to actually partition the individual operators in the network. Some operators, such as embedding tables, can be sharded width-wise with minimal overheads. Others, such as matrix multiplies, can still be partitioned (e.g. using parallel GEMM strategies~\cite{BLIS1}) but involve more communication steps. Figure~\ref{fig:embedding_table_parallel} illustrates the example of a model-parallelized embedding table. These width-wise sharding strategies, more generally known as \textit{tensor parallelism} as they require input tensor partitioning, can enable more performant intra-operator parallelism versus inter-operator model parallelism, but require more effort and thought to implement. In addition, the performance benefits are tempered by the fact that most tensor parallel operators require at least one all-gather communication step to reaggregate partitioned outputs. Mesh-TensorFlow~\cite{shazeer2018mesh} provides a general framework and language for specifying tensor-parallel computation, though it should be noted that it is no longer actively supported or maintained.

Model parallelism of any sort introduces GPU-GPU communication. The latest Nvidia GPUs support ``NVLink'' interconnects --- high-speed GPU-GPU communication routes that offer as much as 900GB/s bandwidth --- which can help minimize overheads. NVLink is not always readily available, however, particularly when using cloud-provided machines that the user cannot customize easily. When NVLink is not supported, GPU-GPU communication runs over PCIe interconnects, which are much slower. The Tesla V100, generally considered a standard high-performance GPU for DL applications, supports a 16-lane PCIe 3.0 interconnect, at 16GB/s. 

To avoid having to transfer too much data over slow interconnects, model parallelism users generally aim to select a partitioning strategy that will minimize the size of activations that need to be transferred between shards, or else balance out computation to hide communication costs. Various sharding algorithms exist for this~\cite{flexflow2018,gpipe2018,lamp2020,mpanalysis2019,hydra2021,zero2019}.

\textbf{Data parallelism} is a common deep learning execution strategy that enables multiple mini-batches of data to be consumed in parallel. Data parallel execution techniques can be divided into two broad categories --- asynchronous data parallelism and synchronous data parallelism. 

The most well-known asynchronous technique is Parameter Server, wherein one core chief server holds a baseline set of parameters while distributed workers hold model replicas that train on different mini-batches. The distributed workers occasionally send updates to the baseline server, which in turn will send out replacement parameters to the distributed workers to keep them updated. The workers can run out of sync with each other, as they only need to communicate/synchronize with the baseline server. Asynchronous techniques introduce many challenges, such as accuracy degradation versus single-worker training and irreproducible results due to variance in worker return times. For these reasons, asynchronous techniques are generally being phased out in favor of synchronous techniques in the modern DL training landscape. 

The most popular synchronous data parallel execution technique is Distributed Data Parallelism (DDP). DDP replicates a model and assigns copies to $m$ different accelerators. An initial ``global mini-batch'' is taken in, then partitioned evenly across the replicas to to produce local gradient updates for each replica. These gradients are then aggregated across replicas to produce a global update, typically using an all-reduce communication pattern. This global update is then applied to every replica in parallel. This technique is mathematically equivalent to single GPU training with the original global mini-batch. While this technique induces an all-reduce communication step, these overheads can generally be overlapped and hidden under model execution times.

\textbf{All-Gather and All-Reduce} communication patterns are often used in data parallelism and broader distributed DL. The aim of these patterns is to take separate data on different processors then aggregate and distribute them back over the processors such that every processor now holds replicas of the same data. All-gather patterns use an algorithm wherein every processor communicates its data to every other processor. This is generally used if each processor holds \textit{a partition} of data that needs to be broadcasted globally. Typically this requires $n$ communication steps per processor with heavy bandwidth usage --- each processor must communicate to all other processors. All-reduce patterns are a layer on top of all-gather that combines aggregation with some reduction function (e.g. sum, average). Running the function \textit{during} synchronization eliminates a step versus running an all-gather followed by a local function application. Horovod~\cite{horovod2018}, for example, implements a bandwidth-optimal reduce pattern~\cite{bandwidth2008} for data parallel gradient aggregation that requires $2 \times (n-1)$ communication steps per processor to complete the full gather and reduce.

\begin{figure}[th!]
\centering
	\includegraphics[keepaspectratio=true, width=0.9\linewidth]{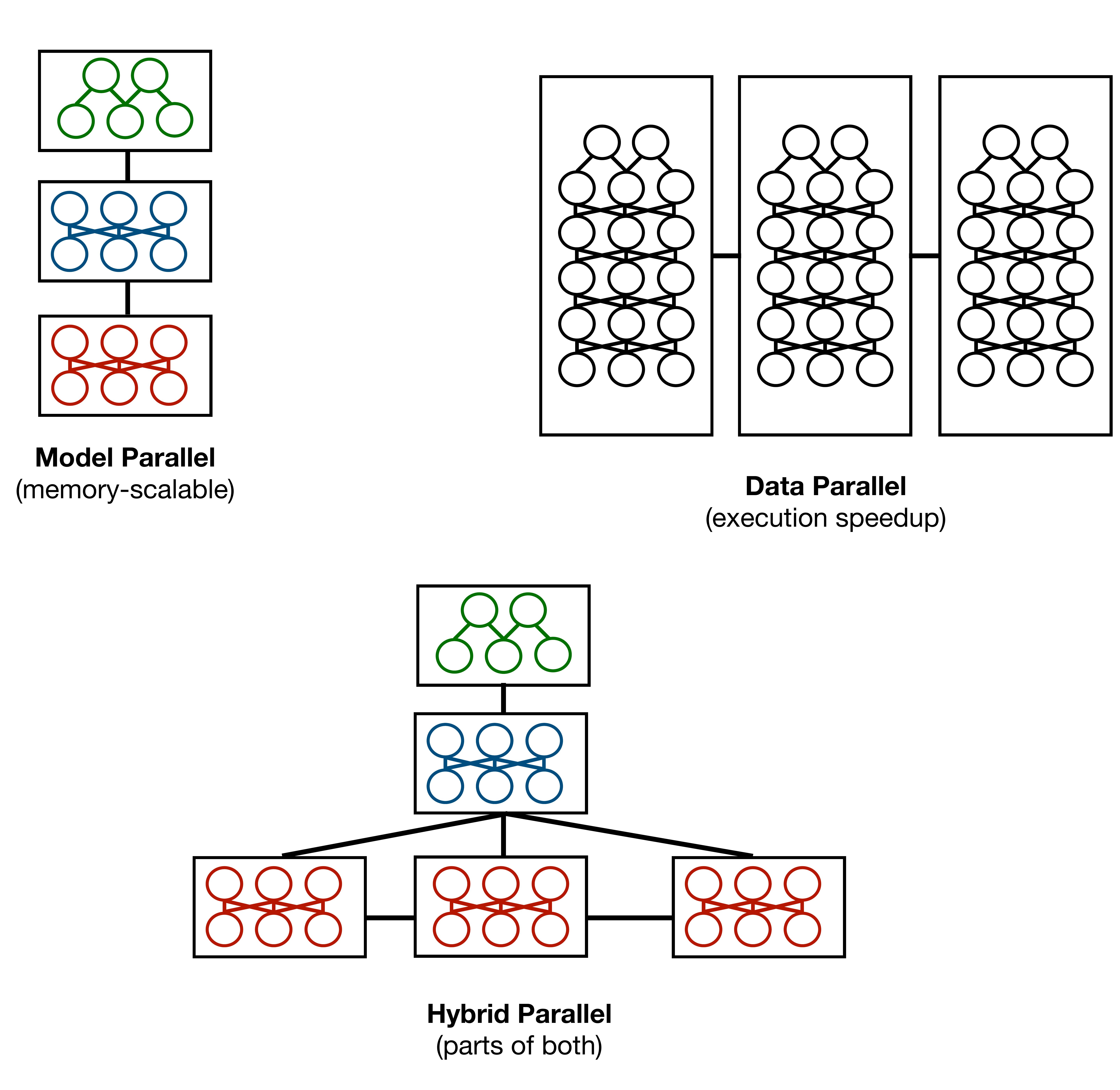}
	\caption{Contrastive illustrations of model parallelism, data parallelism, and hybrid parallelism. Model parallelism enables large-model training across the memory of multiple GPUs, but the speedup potential is heavily reliant on the architecture. Data parallelism offers an easy way to speed up execution generically applicable to most architectures. Hybrid parallelism offers a way to combine the two, but opens up new challenges in finding optimal hybridization strategies and resource apportionment.}
	\label{fig:hybrid_parallel_comparison}
\end{figure}

\textbf{Hybrid parallelism} refers to strategies which \textit{combine} different parallelization strategies to achieve higher overall performance. For example, overlaying data parallelism on top of model parallelism could enable a user to achieve memory scalability across multiple devices along with the execution speedups of data parallelism. These strategies have tradeoffs that need to be factored into their design. In a simple overlay hybridization, model parallelism's multi-device requirements are multiplied by the replication requirements of data parallelism. A further overlay of task parallelism (e.g. in multi-model training) could add another multiplication factor into the equation. More complex hybridizations might apply data parallelism to some \textit{stages} of a model parallel architecture, letting others execute in series, as illustrated in Figure~\ref{fig:hybrid_parallel_comparison}. Note the new ``search space'' this design opens up --- which stages should be chosen for data parallel replication? How do model parallel sharding decisions affect data parallel performance? How should limited resources be apportioned to stages, and how do device interconnects and topologies affect performance? Various tools for hybrid parallel strategy selection designed to answer these questions are covered in Section~\ref{sec:hybrid_parallel_strategy_finding}. Alternative hybrid-parallel designs that attempt to propose \textit{new forms} of parallel execution hybridizing the basic model-, data-, task- parallel approaches are described in Section~\ref{sec:mt_parallel}.

\vspace{-2mm}
\section{Large Model Architectures}\label{sec:large_model}
To clarify the need for large model training systems and motivate their design, we describe various large model architectures and the unique challenges each presents. We can broadly classify the scale of model architectures under two categories --- depth-wise scaling and width-wise scaling. Depth-wise scaling is most commonly needed for long, sequential chain architectures like Transformers. Width-wise scaling is commonly used for very wide, easily parallelized operators (e.g. table lookups). A third setting, example scaling, describes the setting where the size of input samples drives the scaling challenge rather than the size of the model.This setting does not fall under the scope of ``large-model training'', so we do not discuss it in-depth in this paper.

\subsection{Deep Models \& Transformers}
Transformers, the most common example of very deep model architectures, have recently become very popular in a variety of domains. Benchmark tasks in natural language processing (NLP), computer vision (CV), and even tabular data analysis are now led by Transformer architectures. These models consist of stacked ``self-attention'' blocks, each of which consists of multiple dot product operations and matrix operations. Practitioners have found that increasing the depth of Transformers by stacking on more attention blocks generally improves accuracy~\cite{transformerStack}. As such, Transformers have motivated practitioners to explore increasingly deep model architectures. 

These models present a critical challenge for training. One of the first large Transformers was BERT-Large~\cite{bert2018}, using 345M parameters. The memory demands of training this model with a reasonable mini-batch size on a GPU already required practitioners to use high-end GPUs. Since then, Transformers have only become deeper and deeper --- some recent ones have reached one trillion parameters, demanding space well beyond the memory capacity of \textit{any} GPU on the market. 

As such, techniques such as model parallelism are essentially necessary for large Transformer training, and deep model training in general. However, enabling parallel execution for very deep models can be challenging. The most natural sharding strategy for a deep sequence of layers is to partition the sequence into subsequences. But this approach forces the user to add GPUs without actually benefiting from any performance benefits ---- partitioning a sequence into subsequences does not offer any opportunities for parallel execution speedups. Consider a trillion parameter model that needs to use 1024 GPUs to even fit in memory. All of these GPUs are only being used to ``enable'' execution, and provide no performance benefits. In fact, the strategy would likely be slower than an equivalent in-memory training job due to the inter-GPU communication costs.

Some strategies for width-wise sharding exist, such as parallelizing operations in attention blocks across multiple GPUs. However, these approaches require more customization, add communication overheads, and require substantial effort on the part of the model designer to implement. As such, most systems for deep model training prefer to apply a generalized depth-wise sharding strategy that can be optimized for all deep model classes rather than targeting a single architecture at a time.

Despite the challenge of sequential dependencies, depthwise-sharding can introduce many opportunities as well. Techniques such as \textit{pipeline parallelism} and \textit{spilling} only work on depth-wise sharded models. We expand more on these techniques in Section~\ref{sec:mlsys}.

\subsection{Wide Models \& Embedding Tables}~\label{sec:embedding}
Wide models present a very different set of challenges from deep models. While it is easier to parallelize them in a performance effective manner (widthwise rather than depthwise), widthwise partitioning generally requires all-gather communication steps to aggregate parallelized partial outputs.

 \textit{Embedding tables} in recommender models are generally the most popular candidates for width-wise sharding. Embedding-based recommender models are used at most companies which gather entity-specific data (e.g. Meta, Netflix, TikTok) to create customized experiences. A standard approach is to create a table that maps user IDs to trainable vectors that can then be fed to some other DNN placed on top. However, for this to work on a multi-billion user platform such as Facebook, the corresponding table has to be very wide. A three billion index table with size 1024 trainable vectors filled with single-precision (32-bit) floats would require 12TB of memory. A real-world recommender might include \textit{multiple} such tables for different lookups (e.g. user table, business table, video catalog table), further increasing memory costs. 

Partitioning an embedding table is an easy task, given that table lookups are embarrassingly parallel --- a lookup on one index does not rely on other indices. As such, sharding a table into subtables assigned to different GPUs is a common strategy to distribute memory costs. Parallel execution across shards can easily be achieved by simply routing index lookup requests in a mini-batch to the appropriate GPU. However, in order to re-aggregate the mini-batch after the parallel table lookups to feed to the top DNN, a potentially expensive all-gather communication step is necessary.

It is less common, but not unheard of, to apply width-wise sharding to other operators such as matrix multiplies. But in general, embedding tables are the most memory-intensive single operators~\cite{dlrmscale2020}. Given that the primary use-case for width-wise sharding is embedding tables, optimizing for this case may seem overly specific. However, embedding tables and recommender models make up an outsized proportion of DL workloads --- Meta reports that 50\% of their DL training cycles are spent on embedding-table-based recommender models~\cite{dlrmscale2020}. As such, optimizing the very wide model case is well worth the effort even if the applicability is more limited than optimizing for sequential deep model scalability.
\vspace{-2mm}
\section{Large Model Training Systems}\label{sec:mlsys}
\begin{figure*}[th!]
\centering
	\includegraphics[keepaspectratio=true, width=0.7\linewidth]{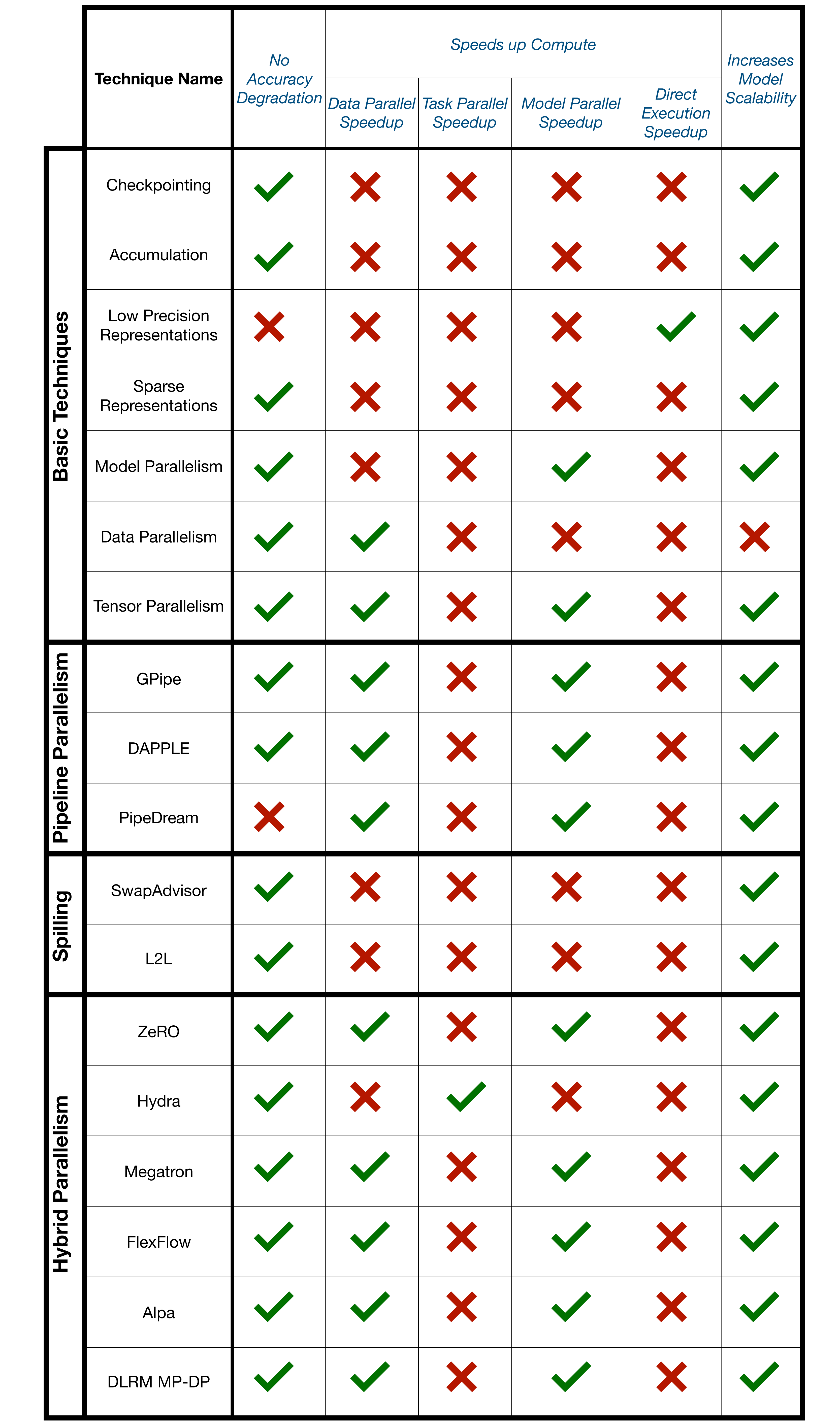}
	\caption{A full comparison of various large model training systems and techniques.}
	\label{fig:technique_comparisons}
\end{figure*}

Several systems have emerged to address the challenge of enabling efficient large-model training. We now describe the major classes of large-model training systems as well as key instances of each category. Figure~\ref{fig:technique_comparisons} provides a tabular comparison of the various systems we cover in this survey.

\subsection{Basic Techniques}
A few basic techniques such as rematerialization are often used as common building blocks for more advanced large-model training systems. In general, these techniques have minimal impact on organization and structure, making them amenable for integration with other approaches.

\subsubsection{Rematerialization}
Rematerialization, also known as gradient checkpointing, attempts to minimize the memory demands of \textit{backpropagation} specifically~\cite{checkpointing2000, checkpointing2016}. As explained in Section~\ref{sec:background}, backpropagation requires saving intermediate operator outputs for proper application of the chain rule for gradient computation. However, intermediate output tensors can demand a great deal of memory!  Some analyses~\cite{lowmemory2019} have shown that activations make up as much as 95\% of memory consumption for ResNet~\cite{resnet2015} and as much as 80\% of memory usage for some Transformers. Rematerialization trades compute for memory by initially discarding most of the activations except for a few \textit{checkpoints}, then \textit{recomputing} the discarded during backpropagation using the checkpoints. In this way, only the intermediates between checkpoints need to be stored in memory at any given point. This approach does induce computational overhead --- the forward pass is effectively being run \textit{twice}. However, the operators in the forward pass are generally faster than the automatic differentiation procedure used in backpropagation, so the overhead is smaller than it might seem. Some gradient checkpointing systems claim to have only 30\% overheads for 6-7X memory savings~\cite{checkpointing2016}. Checkpointing is critical to techniques such as \textit{pipeline parallelism} and \textit{shard alternator parallelism}, described in sections \ref{sec:pp} and \ref{sec:mt_parallel}.

\subsubsection{Accumulation}\label{sec:accum}
Accumulation targets the memory demands of batched \textit{gradients} in backpropagation~\cite{gpipe2018}. Section~\ref{sec:background} describes how stochastic gradient descent batches samples into mini-batches that are fed through the model. In turn, we can consider the gradients that are produced for parameter updates to be the aggregation of the updates that would have been applied for each sample. Accumulation delays the application of these aggregated gradients, instead computing new mini-batch-gradient-updates and accumulating them onto our aggregated gradient vectors. The new gradient is now the aggregated sum of \textit{2} mini-batch updates, rather than 1. In this way, we can scale up our effective mini-batch size and gradient impact without actually training a larger batch. We refer to the smaller, individual batches as \textit{microbatches}, and keep referring to the effective summed batch as the mini-batch. Accumulation is essential to \textit{pipeline parallelism}, and is often used in conjunction with other techniques.

\subsubsection{Low Precision Representations}
Most training frameworks (e.g. TensorFlow, PyTorch)\cite{tfpaper, torchpaper} use single-precision float (32 bit) representations of gradients and parameters. Double-precision representations (64-bit) are relatively uncommon. One way to reduce the memory demands of training a model is to use \textit{half-precision} (16 bit) representations of data. Naturally, this induces an accuracy loss as values are being approximated~\cite{nakandala2020incremental}. However, this approach can offer both speedups and memory savings. To try and balance this, \textit{automatic mixed precision} (AMP) ~\cite{amp2020} will automatically try and determine when data can be safely compressed to 16 bit without accuracy losses. AMP generally reports little-to-no accuracy losses while achieving as much as 5.5X speedups when training large models~\cite{amp2020}. Since AMP is directly modifying values at a very low-level, this technique is generally orthogonal to actual systems approaches for large-model training. 

\subsubsection{Sparse Representations}
In some cases, the vectors used in DL training are very sparse. As an example, embedding table lookups generally only involve a few indices of the table. The gradient vector applied to the table will only have non-zero values at the used indices, while the rest of the gradient will be zeroed out. Actually maintaining all of these zeroes in memory is unnecessary and wastes memory. Sparse representations attempt to compress these vectors down to their non-zero values while avoiding any information loss. The most simple approach, commonly used by default for embedding tables, is to represent a gradient as key-value pairs mapping indices to gradient values. An example is illustrated below.

\[ <0, 0, 0, 0, 0, 0, 4, 0> \rightarrow \{6: 4\} \]

This representation can easily be mapped back for application while discarding unnecessary zero values.

Some complications arise when combining sparse representations with operations that assume standard vector representations, such as all-reduce communication patterns. Some works~\cite{mlplatformmeetup2022} show how this can be resolved with alternative communication patterns or by converting data back to standard representations. Sparse vector representations address a very specific problem, but are critical for efficient training of some operators such as wide embedding tables.

\subsection{Pipeline Parallelism}\label{sec:pp}

Pipeline parallelism targets the ``sequential deep model'' setting~\cite{gpipe2018}. It is a direct extension of the model parallel training paradigm described in Section~\ref{sec:background}. Model parallelism creates a staged-out sequence of shards, creating a natural ``pipe'' structure. Pipelining simply exploits this pipe structure by attempting to fill up the stages with execution operations, reducing the idling present in pure sequential model parallelism suffers. Consider the example of a pure feedforward network, like the one illustrated in Figure~\ref{fig:model_parallel_combined}A). Each shard can be considered a stage of the pipe, such that a model partitioned three-ways over three GPUs is now a three-stage pipeline.

\subsubsection{Pipelining Basics}

\begin{figure}[th!]
\centering
	\includegraphics[keepaspectratio=true, width=\linewidth]{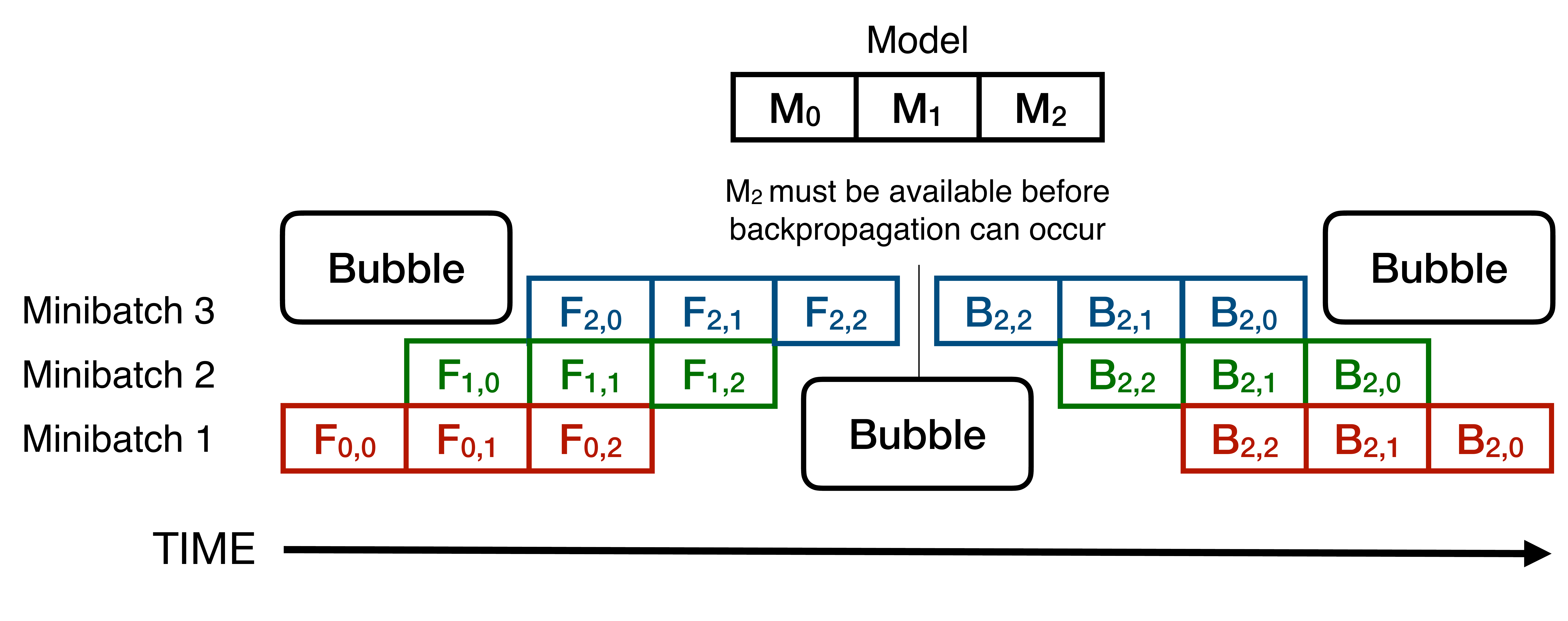}
	\caption{An illustration of pipeline parallelism over three shards. The input mini-batch is partitioned into \textit{microbatches} that are then shuttled through the pipe stages. $F_{x,y}$ refers to the forward stage on shard $x$ with mini-batch $y$, while $B_{x,y}$ 
	refers to the backward stage on shard $x$, mini-batch $y$. In this way, a sort of ``pipelined'' data parallelism is achieved, where data is processed in parallel \textit{across different model parallel stages}. Note that before backpropagation, the forward pipeline must be cleared.} 
	\label{fig:pipeline_parallel}
\end{figure}

In CPU pipelining, we fill up the pipeline with various instructions being sent to the CPU\cite{shen2013modern}. For DL pipelining, we fill up the pipeline with microbatches, like those used in gradient accumulation~\cite{gpipe2018,torchgpipe2020}. In essence, pipeline parallelism is the combination of gradient accumulation and model parallelism. Independent microbatches are shuttled through the shard pipeline, then gradients for each microbatch are accumulated for each pipeline stage. Once the gradients for the full mini-batch (combination of all microbatches) are all aggregated, they can be applied to the model. This design is almost like a model-parallel-data-parallel hybrid, where data shards are being processed in parallel, but on different model parallel shards. Figure~\ref{fig:pipeline_parallel} illustrates.

\subsubsection{Challenges}
Backpropagation presents a challenge for pipeline parallel training. As explained in Section~\ref{sec:background}, intermediate outputs must be available for backpropagation to occur. When combined with accumulation, however, this would require us to store a different intermediate output set for each microbatch, thus robbing us of any scalability advantage offered by accumulation. GPipe~\cite{gpipe2018}, one of the first pipeline parallel training systems, proposed combining accumulation with checkpointing to address this issue. Activations would only be stored at shard/pipe stage boundaries, with recomputation occurring as gradients shifted backwards through the pipe during backpropagation. The checkpointing approach is now standard in most, if not all, pipeline parallel training systems~\cite{pipedream2018}.

Another challenge is presented by the structure of the pipe. For pipelining to work, the shard pipe must be bidirectional. Input and activations flow forward during prediction, and gradients flow backward during backpropagation. This leads to a problem --- data within the pipe will ``collide'' on stages as it flows in both directions. As such, a pipeline flush occurs between prediction and backpropagation. The flush can severely hurt performance if not properly managed. Figure~\ref{fig:pipeline_parallel} illustrates a pipeline parallelized model. Note that a significant portion of time is spent on ``bubble'' periods, where the pipeline must be flushed out fully.

\subsubsection{Major Pipelining Systems \& Techniques}

There have been many attempts to address the aforementioned challenges. GPipe~\cite{gpipe2018} suggested increasing the number of microbatches while keeping accelerator counts constant, so the pipe could stay full for longer. This would not \textit{eliminate} the flush, but it would improve overall efficiency. However, this approach would demand more memory to store more checkpointed microbatch activations. DAPPLE~\cite{dapple2020} proposed an alternative pipelining schedule that could maintain GPipe's convergence behaviors but with less idling. Unfortunately, it also increases memory costs substantially by keeping more microbatches ``alive'' at once, making the schedule infeasible for applications already pushing the boundaries of memory.

Another solution was proposed in the form of \textit{asynchronous pipelining}, which would reorder pipeline stages and backpropagation to eliminate the flush, at the cost of maintaining strict convergence behaviors. This ``decoupling''  of the order relaxes the problem into a more efficient one --- at the cost of affecting data consumption ordering and consumption~\cite{pipedream2018,li2021chimera,narayanan2021memoryefficient,yang2020pipemare}. For example, the 1F1B pattern proposed by PipeDream~\cite{pipedream2018} runs one forward stage for every backward stage (on different microbatches) to maintain a perfect ratio and utilization. But its design requires more careful partitioning and packing, and mitigating accuracy degradation from stale weight updates requires stashing multiple copies of weights~\cite{pipedream2018}, thus blowing up memory costs. While asynchronous pipelining like 1F1B can perform well, it is not a general solution --- the accuracy losses are case-specific and can often be substantial. Applications where accuracy is critical and convergence behaviors must be replicable (e.g. model selection) are not a good fit for asynchronous pipelining.

\subsection{Memory Offloading \& Spilling}\label{sec:spilling}
While model parallelism looks at execution over multiple GPUs to distribute memory demands, some systems attempt to make use of main system memory (DRAM) rather than horizontally scaling across more GPUs. The primary motivation for this approach is that while GPU memory is limited and expensive, DRAM is substantially cheaper and accessible.

\subsubsection{Motivation}
Consider an AWS-provided p3.2xlarge node. It has a single V100 GPU with 16GB of on-device memory. A large DL model might not fit into the node's GPU memory, but in actuality the node has far more \textit{total} memory still available --- 61GB of DRAM plus the 16GB of GPU memory. DRAM (sometimes referred to as CPU memory in the ML systems literature~\footnote{We consider the term CPU memory is ambiguous, given that it could be interpreted to refer to registers or CPU caches. In this paper, we use the term DRAM to refer to main system memory, but it should be noted that other ML systems papers may use the phrase CPU memory to refer to the same concept.} is far cheaper and more available than GPU memory, and a standard cloud-provided multi-GPU machine might easily have hundreds of GBs of DRAM. 

If a model's memory demands can be spread across both DRAM and GPU memory, a large model could be trained without the need for model parallelism's multi-GPU costs. A on-demand AWS p3.2xlarge node offers 77GB aggregate memory at a rate of \$3.06 per hour. In order to have the same amount of GPU-only memory, a p3.16xlarge 8-GPU node would be necessary --- costing the user \textit{8X as much} at \$24.48 per hour.  The cost benefit is clear, and as such, offloading part of the memory demands of a model to DRAM rather than scaling across multiple GPUs can be an attractive option for cost-conscious users with limited GPU access (e.g. small enterprises and researchers). 

\subsubsection{Offloading Systems \& Techniques}

Many initial works~\cite{tflms2019,meng2017training,swapadvisor2021,vdnn2016,wang2018} treated offloading as a ``swapping'' problem --- deciding when to swap tensors off of GPU memory and onto DRAM. Most use graph analysis algorithms to determine where to ``inject'' a swap operation based on when an activation, gradient, or parameter might next be used in the execution graph. SwapAdvisor, the most advanced of these swapping systems, uses a parallelized genetic search algorithm to analyze where the swap operators should be placed for best performance. It was also one of the first systems to support offloading \textit{parameters} as well as activations, which is critical for training billion-parameter model architectures.

These complex swapping procedures can be difficult to setup --- SwapAdvisor's search algorithm takes roughly an hour to complete. Moreover, they are difficult to extend to multi-GPU training, as there is no clear way to extend the swap-injected graph technique to cover multi-GPU parallelism.

Another approach was proposed with ZeRO-R~\cite{zero2019}, a system for offloading that sends activations and parameters to DRAM dynamically. This approach ``offloads when needed'', rather than planning offloads up front. The irregularity of the design can introduce issues such as memory fragmentation, but it adds a great deal of flexibility versus graph-based designs. A later version, ZeRO-Infinity~\cite{zero2021} extended this to offloading to NVMe/disk storage for further scalability.

Hydra~\cite{hydra2021} opts for an ``independent block'' strategy, dividing a model architecture into submodels (like model parallelism) that can then be spilled between DRAM and GPU memory freely. An analogy can be drawn to spilling in RDBMSs, where independent data chunks can be sent down to a lower level of memory. Unlike other spilling systems, Hydra's execution pattern is identical to model parallelism, and separates the execution of each model shard entirely. It still tries to overlap communication and computation, but ignores the complexities of fine-grained tensor offloading explored by other CPU-offloading techniques. This generalization makes it less-than-optimal for single-GPU execution, but makes it far more amenable to hybridization with multi-GPU parallelization techniques. We expand on this further in Section~\ref{sec:mt_parallel}.

\begin{figure}[h!]
\centering
	\includegraphics[keepaspectratio=true, width=\linewidth]{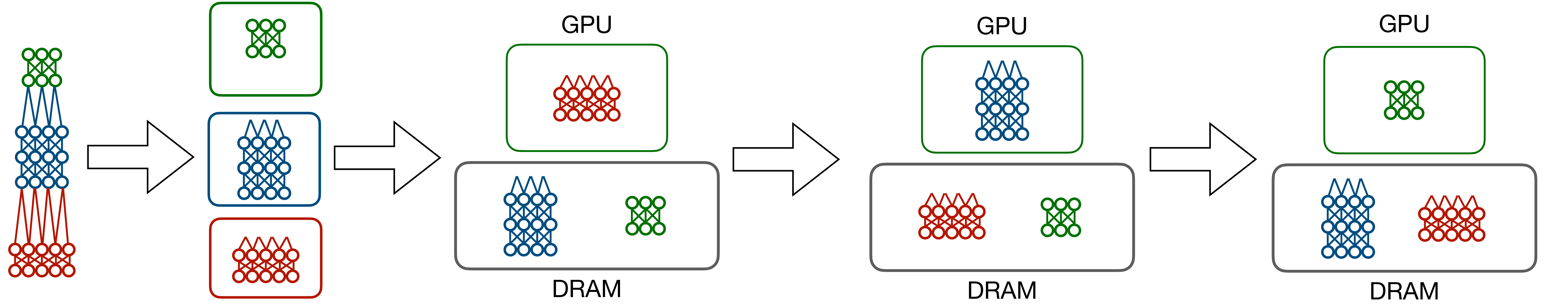}
	\caption{Hydra's spilling strategy simply promotes and demotes model parallel shards on and off of GPU memory. Other spilling designs such as the one used by ZeRO-Offload are similar, though less strictly structured.}
	\label{fig:shard_parallel}
\end{figure}

L2L~\cite{l2l2020} uses a design similar to Hydra's but is more restricted in its sharding approach. It targets Transformer architectures specifically, and swaps self-attention blocks (standard Transformer operators) with heuristics selected specifically for its target class of models. This allows it to perform very well on Transformer architectures, but prevents it from achieving the flexibility of Hydra or the dynamic generality of ZeRO-R.

Note that these techniques are generally used for distributing \textit{depth-wise} large model memory demands, as they all exploit some kind of sequential ordering in execution. A very wide operator (e.g. an embedding table) that cannot be serialized without substantial performance slowdowns, cannot easily be spilled across DRAM and GPU memory. The only option for hybrid-device execution on wide operators is to either serialize the parallel operator (index lookup in the table case) and rewrite the series of operations into a deep, rather than wide, model, or else to actually execute the wide operator on the CPU. 

\subsubsection{Offloaded Computation}
Some systems go further still, actually executing operations on the CPU. In general, it is preferable to run a model entirely using GPU or TPU compute, as most DL operators will run much faster on accelerators that support high degrees of parallelism. With offloading, however, the data will be on the CPU anyway --- so executing CPU operations in parallel with GPU operations should not add overheads.

ZeRO~\cite{zerooffload2021} proposed running parameter updates on the CPU while GPU execution is ongoing, specifically for the popular Adam optimizer~\cite{adam2014}. The Adam optimizer holds some state parameters (typically 32-bit) and needs to run on 32-bit parameters to avoid accuracy degradation. Unfortunately, this prevents users from exploiting 16-bit representations for reduced memory demands. The ZeRO version of the Adam optimizer maintains 32-bit versions of the parameters on DRAM and low-precision 16-bit versions on the GPU to consume less memory. During execution, the system spills gradients and optimizer state onto DRAM, then runs parameter updates on the 32-bit parameters using \textit{CPU processing}. The updates are propagated back to the 16-bit parameters in a secondary step that overlaps CPU-GPU communication with GPU computation. 

Mixed-CPU-GPU compute is also common for very large recommender models. As explained Section~\ref{sec:embedding}, embedding tables are very wide memory-intensive operators which generally feed into some smaller DNN for further processing. Without any optimization, the sheer scale of the embedding table would force CPU-only execution~\cite{dlrmscale2020}. Alternatively, a user could place the embedding table on the CPU while the DNN sits in GPU memory and enjoys the benefits of GPU acceleration. Some works such as Hotline~\cite{hotline2022} try and pipeline data through the model, from the CPU-based embedding table into the GPU-accelerated DNN. They demonstrate that this mixed compute approach can be even faster than width-wise multi-GPU model parallelism, as it eliminates the need for the all-to-all communication step described in Section~\ref{sec:embedding}.

\subsection{Hybrid Parallelism}

\begin{figure*}[th!]
\centering
	\includegraphics[keepaspectratio=true, width=\linewidth]{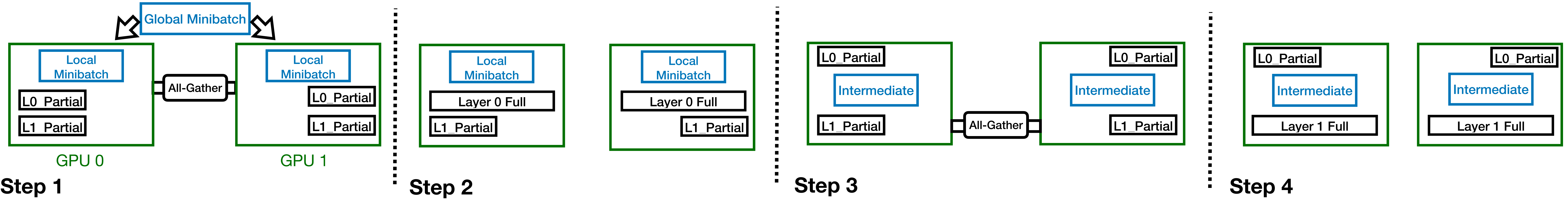}
	\caption{A snippet of FSDP-style of execution on a simple 2-layer model parallelized across 2 GPUs. All-gathers occur at every stage to enable data parallel execution on the current layer, while inactive layers are still partitioned across devices. Thus the benefits of data parallelism and model parallelism are partially merged.}
	\label{fig:shard_parallel}
\end{figure*}

The basic parallelization techniques described in Section~\ref{sec:parallelization} can be combined in different ways. Various systems have attempted to combine the benefits of the various ``basic'' parallel execution approaches (e.g. data parallelism, model parallelism) to offer users higher performance and scalability. Hybrid parallelism techniques can be classified into two broad categories --- ``true'' hybrids that integrate parallelization techniques from the ground up, and top-down hybrids that select between different strategies at different stages of execution. 

\subsubsection{Ground-Up Hybrids}\label{sec:mt_parallel}
Traditionally, combining model parallelism with other techniques from the ground up has been a challenging task. Model parallelism drives up the GPU requirements of standard execution, which can make combinations with replication-based or multi-instance techniques for parallelism (e.g. data parallelism, task parallelism) impractical as they scale up model parallelism's device requirements further still. 

To address this problem, Hydra~\cite{hydra2021} proposed using a spilling technique like those outlined in Section~\ref{sec:spilling} to reduce the number of GPUs needed for scalable model-parallel training, and then applying a layer of task parallelism on top to support efficient multi-model training. The Hydra system then exploits the segmented nature of model parallelism to enable hybrid ``fine-grained parallel'' schedules that can outperform both standard task parallelism and model parallelism. Figure~\ref{fig:shard_parallel} illustrates. At the moment, Hydra is the only system to explicitly target the multi-model setting for very large models, but this area will likely grow in importance as practitioners struggle with the costs of model selection and multi-user cluster management.

\begin{figure}[h!]
\centering
	\includegraphics[keepaspectratio=true, width=\linewidth]{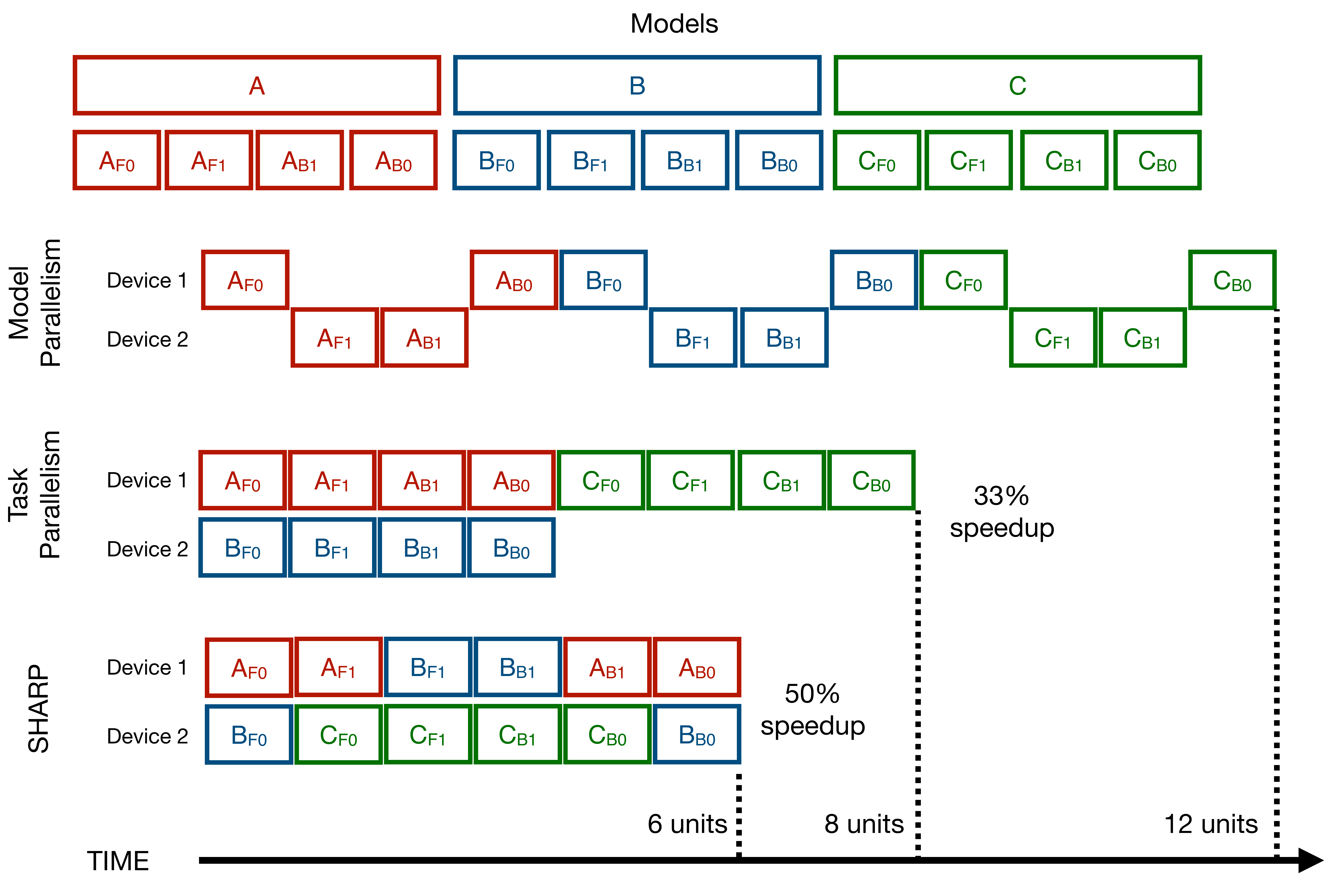}
	\caption{``Shard Alternator Parallelism'' combines sequential model parallel spilling and task parallelism to outperform both standard model and task parallel techniques.}
	\label{fig:shard_parallel}
\end{figure}

Fully Sharded Data Parallelism (FSDP), originally introduced with ZeRO~\cite{zero2019}, offers a hybrid of model parallelism and data parallelism. Unlike Hydra, which still executes in a model-parallel-sharded fashion, FSDP only uses model parallelism to \textit{distribute} the model over data parallel instances with each data parallel clone holding a partial set of parameters for a layer group. When a layer group is executed, FSDP runs an all-gather step to produce the full, unsharded layer group on every data parallel instance. The layer group is then executed in a purely data parallel fashion. After the layer is executed, it can be resharded immediately after to redistribute the memory footprint. A similar approach is used for backpropagation.

Under FSDP, the per-accelerator memory requirements are reduced down to a minimum footprint of a single layer plus the partitioned memory requirements of the rest of the layers. Discarding the single layer requirement as a constant factor, we can represent this as an $O(n/k)$ reduction, where $n$ is the original model memory footprint and $k$ is the number of data parallel instances. This enables users to simultaneously benefit from the performance of data parallelism and the scalability of model parallelism. Note that this does add substantial communication overheads --- an all-gather is run for every layer --- and still requires horizontal scaling for scalability, unlike spilling-based techniques.

ZeRO-Offload~\cite{zerooffload2021} proposed combining FSDP with spilling per-accelerator, offloading sharded layer parameters that will not be used in the near future. This offers substantially better scalability, but introduces more communication overheads through CPU-GPU communication. ZeRO works to overlap the communication with compute, but some slowdown is generally inevitable. Analyses have shown that FSDP is slower than standard data parallelism (though more scalable and capable of running substantially larger models). Proponents of FSDP claim that users can exploit its higher scalability to increase batch sizes (and thus bring execution times in line with DDP performance), but we note that batch size can affect accuracy convergence behaviors. Scaling batch size for better FSDP performance can lead to the same issues that we outlined in our discussion of asynchronous pipelining (though less extreme).

3D Parallelism combines FSDP with pipeline parallelism and tensor parallelism to exploit scalable data parallelism along with parallel depth-wise and width-wise sharded execution. This usually takes the form of applying FSDP in some parts of the model, pipelining in another, and tensor parallelism in another segment more amenable to width-wise sharding. 3D parallelism generally requires a great deal of customization based on the model architecture --- it cannot be applied out-of-the-box like Hydra or FSDP. That being said, it has been applied successfully using systems such as Megatron~\cite{megatron2019} in the training of many very large-scale models such as Megatron-LM~\cite{megatronlmblog2020} and BLOOM~\cite{bloom2022}. In the future, a new ``4D parallelism'' might be possible by combining 3D parallel hybrids with Hydra's sharded task parallelism.

\subsubsection{Strategy Finding}\label{sec:hybrid_parallel_strategy_finding}
Strategy-finding systems attempt to automate the process of combining parallelization techniques within a model. A few recent examples are FlexFlow~\cite{flexflow2018} and Alpa~\cite{alpa2022}. 

FlexFlow, which was built prior to the development of advanced DL parallelization techniques such as pipeline parallelism, FSDP, and sharded task parallelism, only explored data, tensor, and model parallelism, and primarily targeted convolutional neural networks. FlexFlow builds a device topology graph that models accelerators as nodes and interconnects (e.g. NVLink, PCIe, Infiniband network) as edges. This allows it to produce hybrid parallel execution strategies that account for the cost of data movement between edges in a given device configuration. It uses a simulator to evaluate different partitioning strategies, using pilot passes to model operator runtimes and theoretical calculations based on edge bandwidths to model communication overheads. Using the simulator as an oracle, it evaluates different ways to partition the operators. Note that this ``partition'' based representation of parallelism cannot support parallelization techniques that exploit independent execution on different tasks (e.g. task parallelism, pipeline parallelism), though it could potentially support FSDP. In addition, it does not explicitly account for memory scalability or the possibility running out of device memory in a particular configuration~\cite{memflow2020}.

Alpa 	accounts for memory scalability more explicitly, and considers inter-operator parallelism (e.g. model parallelism, pipeline parallelism) rather than just intra-operator partitioning like FlexFlow. It uses an ILP formulation to determine how to setup the parallelization strategy, then modifies the execution plan if a stage will exceed device memory limits~\cite{alpa2022}. This approach can enable better performance than FlexFlow by accounting for a broader strategy search space.

These hybrid parallelization strategies are well suited to static, non-data-dependent execution tasks (e.g. non-recurrent neural networks). However, they do not scale well to more dynamic tasks such as multi-model training --- they are compilers for training, not schedulers. Future works could consider bridging this gap, building a dynamic hybrid parallel executor.

\subsubsection{Model-Data Parallelism for Recommender Models}
DLRMs present a unique challenge to practitioners because they combine two different scaling challenges. The embedding tables are very wise, and warrant width-wise partitioning for model parallel execution. The top DNN is computationally intensive but small, and would benefit most from data parallelism.  As such, a hybrid strategy that applies tensor parallelism to the table of the model and data parallelism to the DNN would perform well on recommender models. This approach has become the standard for fully GPU-accelerated DLRM training~\cite{dlrm2019} though the heterogeneous CPU-GPU execution that we described previously is also popular for users with less access to GPU resources. 

\begin{figure}[h!]
\centering
	\includegraphics[keepaspectratio=true, width=\linewidth]{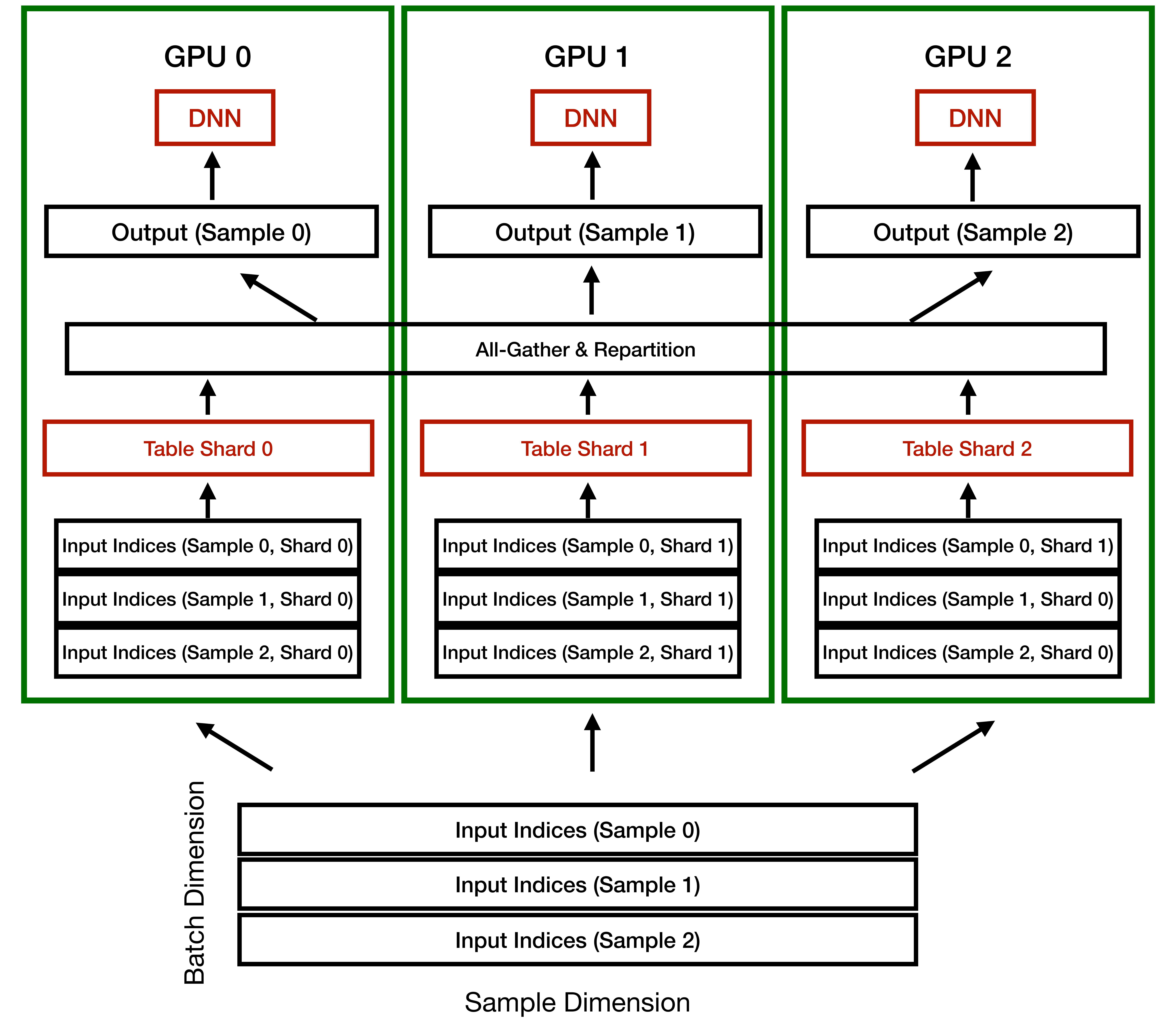}
	\caption{Illustration of the dataflow and execution patterns when training a DLRM using hybrid model-data parallelism across 3 GPUs with a mini-batch size of 3.}
	\label{fig:mp_dp_parallel}
\end{figure}

Hybrid parallel DLRM training partitions the embedding table over multiple GPUs, and places a local copy of the top DNN on each GPU. The sharded table processes an input sharded across the \textit{sample dimension}, then runs a partitioned all-gather to reaggregate the table outputs and partition them across the \textit{batch dimension} for every data parallel copy. Figure~\ref{fig:mp_dp_parallel} illustrates.

This approach allows practitioners to benefit from both data and model parallelism within the neural architecture. The communication step is intensive, and often induces heavy overheads~\cite{mlplatformmeetup2022}, but this is generally outweighed by the benefits of parallel execution.

Overall, hybrid parallelism offers users the ability to efficiently train models by combining the benefits of different parallelization strategies when appropriate. Blended parallel techniques like sharded task parallelism and FSDP combine scalability and efficiency from the ground-up, while strategy finding and DLRM hybrid parallelism can help train model architectures that have mixed demands at different stages of the graph.

\section{Conclusion and Discussion}
\label{sec:conclusion}
In this paper, we surveyed large-model training systems research, covering various approaches ranging from heterogeneous execution to hybrid parallelism. Large-model training systems research is becoming increasingly critical as practitioners push model scales further and further. Efficiency and scalability have become key concerns for DL practitioners in a variety of domains, and the development of these systems is essential to supporting further advancements in DL.

There are several avenues yet to be explored large-model training research. Existing hybrid parallel training systems still only cover a relatively limited search space, with 3D parallel pipeline-data-tensor parallelism being the most complex configuration available to current users. Future hybrid parallel systems can be pushed further to introduce even more dimensions such as task parallelism for multi-model training and CPU spilling, potentially offering even higher scalability and efficiency.

New hardware developments will also drive new systems advances in this space. New CPU-GPU servers are being developed to support high-bandwidth NVLink communication between the CPU and GPU --- this would speed up spilling communication by nearly 60X versus PCIe 3.0x16 interconnects on Tesla GPUs~\cite{hopper2022}. TPUs and other custom hardware with increased memory capacities also offer an attractive option versus memory-limited GPUs. As the limitations and constraints of training hardware change, we will see shifts in the design and aims of large-model training software systems. 

Current systems do not optimize explicitly for serverless execution --- they would work just as well in a fixed on-premise cluster as they would on cloud-provided machines. But serverless execution offers both new opportunities and challenges. Budgeted training, for example, could allow users to specify how much they are willing to spend on a particular experiment, with automatic parallelization systems then directly determining what resources they ought to provision to support efficient training within the cost constraints. Elastic scaling for dynamic large-scale training could offer many exciting opportunities such as autoscaling dimensions of hybrid parallel designs, or automatically evaluated.

The large-model training systems space will continue to develop, driven forward by DL practitioners' need to explore a wider range of architectures. As the compute and memory demands of DL increase, the importance of this space will only continue to grow. The various directions of research surveyed in this paper will serve as the foundations for new upcoming developments designed to meet the changing demands of the space.

\bibliography{main}
\bibliographystyle{abbrv}
\end{document}